\documentclass[prd,preprint,nofootinbib]{revtex4}

\usepackage{graphicx}
\usepackage{amssymb}
\usepackage{amsmath}
\usepackage{color}

\begin{document}
\renewcommand{\thefootnote}{\#\arabic{footnote}}
\newcommand{\rem}[1]{{\bf [#1]}}
\newcommand{\gsim}{ \mathop{}_ {\textstyle \sim}^{\textstyle >} }
\newcommand{\lsim}{ \mathop{}_ {\textstyle \sim}^{\textstyle <} }
\newcommand{\vev}[1]{ \left\langle {#1}  \right\rangle }
\newcommand{\bear}{\begin{array}}  
\newcommand {\eear}{\end{array}}
\newcommand{\bea}{\begin{eqnarray}}   
\newcommand{\eea}{\end{eqnarray}}
\newcommand{\beq}{\begin{equation}}   
\newcommand{\eeq}{\end{equation}}
\newcommand{\bef}{\begin{figure}}  
\newcommand {\eef}{\end{figure}}
\newcommand{\bec}{\begin{center}} 
\newcommand {\eec}{\end{center}}
\newcommand{\non}{\nonumber}  
\newcommand {\eqn}[1]{\beq {#1}\eeq}
\newcommand{\la}{\left\langle}  
\newcommand{\ra}{\right\rangle}
\newcommand{\ds}{\displaystyle}
\newcommand{\red}{\textcolor{red}}
\def\SEC#1{Sec.~\ref{#1}}
\def\FIG#1{Fig.~\ref{#1}}
\def\EQ#1{Eq.~(\ref{#1})}
\def\EQS#1{Eqs.~(\ref{#1})}
\def\lrf#1#2{ \left(\frac{#1}{#2}\right)}
\def\lrfp#1#2#3{ \left(\frac{#1}{#2} \right)^{#3}}
\def\GEV#1{10^{#1}{\rm\,GeV}}
\def\MEV#1{10^{#1}{\rm\,MeV}}
\def\KEV#1{10^{#1}{\rm\,keV}}
\def\REF#1{(\ref{#1})}
\def\lrf#1#2{ \left(\frac{#1}{#2}\right)}
\def\lrfp#1#2#3{ \left(\frac{#1}{#2} \right)^{#3}}
\def\OG#1{ {\cal O}(#1){\rm\,GeV}}

\newcommand{\ah}{A_H}

\title{High-energy Cosmic-Ray Positrons from Hidden-Gauge-Boson Dark Matter}

\author{Chuan-Ren Chen$^{1}$, Fuminobu Takahashi$^{1}$ and T. T. Yanagida$^{1,2}$}

\affiliation{${}^{1}$Institute for the Physics and Mathematics of the Universe,
University of Tokyo, Chiba 277-8568, Japan\\
 ${}^{2}$Department of Physics, University of Tokyo, Tokyo 113-0033,
Japan }


\begin{abstract}
We provide a scenario in which a hidden $U(1)$ gauge boson constitutes 
dark matter of the Universe and decays into the standard-model
particles through a kinetic mixing with an $U(1)_{B-L}$ gauge boson.
Interestingly, our model can naturally account for the steep rise in the positron fraction
recently reported by PAMELA.
Moreover, we find that due to the charge assignment of $U(1)_{B-L}$,
only a small amount of antiprotons are produced in the decay, which is
also consistent with the PAMELA and other observational data.
\end{abstract}

\preprint{IPMU 08-0086}

\maketitle

\section{introduction}
\label{sec:1} 
The presence of the dark matter has been firmly 
established by numerous observations, although we have not yet understood what dark matter is made of.
It is promising that
the PAMELA~\cite{Picozza:2006nm} and Fermi (formerly GLAST)~\cite{FGST}
 satellites in operation may
reveal the nature of dark matter, helping us to pin down the
dark matter particle. 

Recently much attention has been attracted to a scenario
that the dark matter decays into the standard-model (SM) particles~\cite{Takayama:2000uz,
Buchmuller:2007ui,Bertone:2007aw,Ibarra:2007wg,Ishiwata:2008cu,Chen:2008yi,Ibarra:2008kn,Chen:2008dh,Yin:2008bs}, 
since the energetic particles produced  in the decay may account 
for the excesses in the positron fraction reported by PAMELA~\cite{Adriani:2008zr}
or HEAT~\cite{Barwick:1997ig}  and in the diffuse extra 
galactic gamma-ray background observed by EGRET~\cite{Sreekumar:1997un,Strong:2004ry}.
The PAMELA also released data on the antiproton flux~\cite{Adriani:2008zq}, 
suggesting that most of the observed antiprotons are the secondaries produced by interactions 
between the primary cosmic rays and the interstellar medium. 
The suppression in the antiproton flux from the dark matter decay is particularly important
because some decaying dark matter scenarios predict too large antiproton
flux at the solar system~\cite{Ibarra:2007wg}.

We have recently proposed a scenario that a hidden $U(1)$
gauge boson may decay into the SM particles
through its kinetic mixing with the $U(1)_Y$ in the SM gauge group~\cite{Chen:2008yi}.
In order to explain the longevity of the hidden gauge boson,
the kinetic mixing needs to be suppressed down to an extremely small value,
and direct couplings between the hidden and SM sectors must be
negligibly small~\cite{Pospelov:2007mp}. In order to achieve those features we introduced messenger 
fields having a large mass close to the grand unification theory (GUT) scale,
and assumed that the hidden sector is sequestered from the SM sector.

In this letter we present a model in which such a small
coupling between the hidden $U(1)$ gauge boson and the SM particles
are naturally realized in the extra dimension framework. The essential idea is to separate 
the hidden sector from the SM sector in an extra dimension
so that the two
sectors can communicate with each other only through the interactions with another $U(1)_{m}$
gauge field in the bulk, which is assumed to be broken at a high-energy scale.
The branching ratios simply reflect the charges of the SM particles under
the $U(1)_{m}$ gauge group. Interestingly, our model can account for the steep rise in the positron
fraction reported by PAMELA as well as the gamma-ray excess observed by EGRET. 
Moreover, the antiproton flux can be suppressed  enough to be
consistent with the PAMELA observations,  if $U(1)_{m}$ is identified with $U(1)_{B-L}$.

\section{Set-up}
\label{sec:2} 
We introduce an extra dimension with two branes at the boundaries.
Suppose that the hidden gauge sector
is on one brane and the SM particles are on the other brane well separated from
each other. In such a set-up, dangerous direct interactions between the two sectors
are exponentially suppressed, and  the hidden gauge boson becomes
quasi-stable~\footnote{
The interactions are suppressed by $e^{- M_* L}$, where $M_*$ is the five-dimensional Planck scale
and $L$ denotes the size of the extra dimension. For $M_* L \sim 10^2$, the direct couplings
are so small that the hidden gauge boson will
become practically stable in a cosmological time scale.
Then  $M_*$ is roughly equal to $M_P /10 \simeq 10^{17}$\,GeV
which is larger than the GUT scale ($\sim 10^{15}$\,GeV), and so, our analysis in the text is valid.
}.  It is worth noting that its stability is guaranteed by the geometric separation in the extra dimension,
not by a discrete symmetry such as a $R$-parity. 
If there is a $U(1)_{B-L}$ gauge field in the bulk, the hidden $U(1)_H$ gauge field can have
an unsuppressed gauge kinetic mixing with the $U(1)_{B-L}$.  After integrating out the heavy $U(1)_{B-L}$ gauge boson, the effective couplings between the hidden $U(1)_H$ gauge boson $A_H$ and the SM particles are induced, 
which enables $A_H$ to decay into the SM particles. 
The longevity of the hidden gauge boson
is realized by the hierarchy between $B-L$ symmetry breaking scale and the weak scale. 
As we will see below, taking the $B-L$ breaking scale around the
GUT scale~\footnote{
The seesaw mechanism~\cite{seesaw} for neutrino mass generation suggests the mass of the 
right-handed neutrinos at the GUT scale
$\sim 10^{15}$GeV. Recall that the right-handed neutrinos acquire the masses from 
the $U(1)_{B-L}$ gauge symmetry
breaking. Thus, it is quite natural to consider the $B-L$ breaking scale around the 
GUT scale of $\sim 10^{15}$GeV.
} and the mass of the hidden gauge boson of ${\cal O}(100)$\,GeV  naturally
leads to the lifetime of ${\cal O}(10^{26})$ second that is
needed to account for the positron excess.

Let us first consider the kinetic mixing between the hidden $U(1)_H$ and the $U(1)_{B-L}$ gauge symmetries.
The relevant effective interactions in the four dimensions are written as 
\bea
{\cal L}_{(4D)} &=& -\frac{1}{4}F^{(H)}_{\mu\nu}F^{(H) \mu\nu}
             					      -\frac{1}{4}F^{(B)}_{\mu\nu} F^{(B)\mu\nu}
					             +\frac{\lambda}{2}F^{(H)}_{\mu\nu} F^{(B)\mu\nu}\non\\
					             &&+\frac{1}{2}m^{2} A_{H \mu}A_{H}^{\mu} 
					             +\frac{1}{2}M^{2} A_{B \mu}A_{B}^{\mu},
\label{eq:kinetic}
\eea
where $\lambda$ denotes a coefficient of the kinetic mixing of order unity, and
$F^{(H)}$ and $F^{(B)}$ are the field strengths of the $U(1)_H$ and $U(1)_{B-L}$ gauge bosons,
$A_H$ and $A_B$, respectively. We assume that both gauge symmetries are spontaneously 
broken, and therefore $A_H$ and $A_B$ acquire masses $m$ and $M$, respectively.
We also assume that the kinetic mixing is unsuppressed as $\lambda = {\cal O}(0.1)$,
and we take $m = \OG{100}$ and $M = \OG{10^{15}}$ throughout this letter.
We can make the kinetic terms canonical and diagonalize the mass matrix  by
appropriate transformations. The relations between $(A_H, A_B)$
and the mass eigenstates $(A_H^\prime, A_B^\prime)$ are
\bea
A_H &\simeq& A_H^\prime + \lambda\left(1+ \frac{m^2}{M^2} \right) A_B^\prime,\\
A_B &\simeq&A_B^\prime - \lambda \frac{m^2}{M^2} A_H^\prime,
\label{eq:relation}
\eea
where we have approximated $m^2 \ll M^2$ and $\lambda \lsim 0.1$ for simplicity.

The low-energy effective interactions between
the hidden gauge boson $\ah^\prime$ and the SM fermion $\psi_i$ can be extracted 
from the $U(1)_{B-L}$ gauge interactions by using the relation \REF{eq:relation},
\beq
{\cal L_{\rm int}} \;=\;  q_i A_B^\mu \,  \bar{\psi}_i \gamma_\mu \psi_i \supset
-\lambda\, q_i \frac{m^2}{M^2}  A_H^{\prime \mu} \, \bar{\psi}_i \gamma_\mu \psi_i,
\eeq
where $q_i$ denotes the $B-L$ charge of the fermion $\psi_i$. The partial decay width
for the  SM fermion pair is
\bea
\Gamma(\ah \rightarrow \psi_i \bar\psi_i) &\simeq&\lambda^2 \frac{ N_i q_i^2}{12 \pi} \lrfp{m}{M}{4} m,
\label{pGamma}
\eea
where we have neglected the fermion mass, and $N_i$ is the color factor ($3$ for quarks and $1$ for leptons).
Thus the lifetime $\tau$ is given by
\beq
\tau \;\simeq\; \frac{2.5\times 10^{27} {\rm \,sec}}{\lambda^2 \sum_i N_i q_i^2}  \lrfp{m}{100{\rm GeV}}{-5} \lrfp{M}{10^{15}{\rm GeV}}{4},
\eeq
where the sum is taken over those SM fermions of   masses  lighter than $m/2$.

We show the coefficient $N_i q_i^2$ for the quarks and the leptons in Table~\ref{charge}.  
It should be noted that the branching ratios are not sensitive to the mass of $\ah$
and they simply reflect the $B-L$ charge assignment, which makes our analysis very predictive.

\begin{table}[t]
\begin{center}
\begin{tabular}{|c|c|c|}
\hline
&quark &lepton\\ \hline
$N_i(B-L)^2$& $\frac{1}{3}$ & 1 \\\hline
\end{tabular}
\end{center}
\caption{{The coefficients appearing in the partial decay rates for quarks and leptons.}\label{charge}}
\end{table}%

\section{Cosmic-ray spectra}
\label{sec:3} 
In this section we show the predicted spectra for the
positron fraction, gamma-ray and antiproton fluxes based on the decay modes
shown in the previous section.  More precisely, the branching ratios are $2/39(2/37)$,  $2/13(6/37)$
and $1/13(3/37)$ for a quark pair, a charged lepton pair and a light neutrino pair, respectively, if the top quark decay channel is (not) allowed kinematically.
To estimate the spectra of gamma, positron and antiproton, we use the PYTHIA\ \cite{Sjostrand:2006za} Monte Carlo program. After cosmic-ray particles are produced during the decay of $A_H$, the following calculations 
are straightforward and identical to those adopted in Ref.~\cite{Chen:2008yi},
and so, we show only the final results in this letter. For readers who are interested in 
the details of the calculations should be referred to Ref.~\cite{Ibarra:2007wg} and references
therein.

\begin{figure}[t]
\includegraphics[scale=0.3]{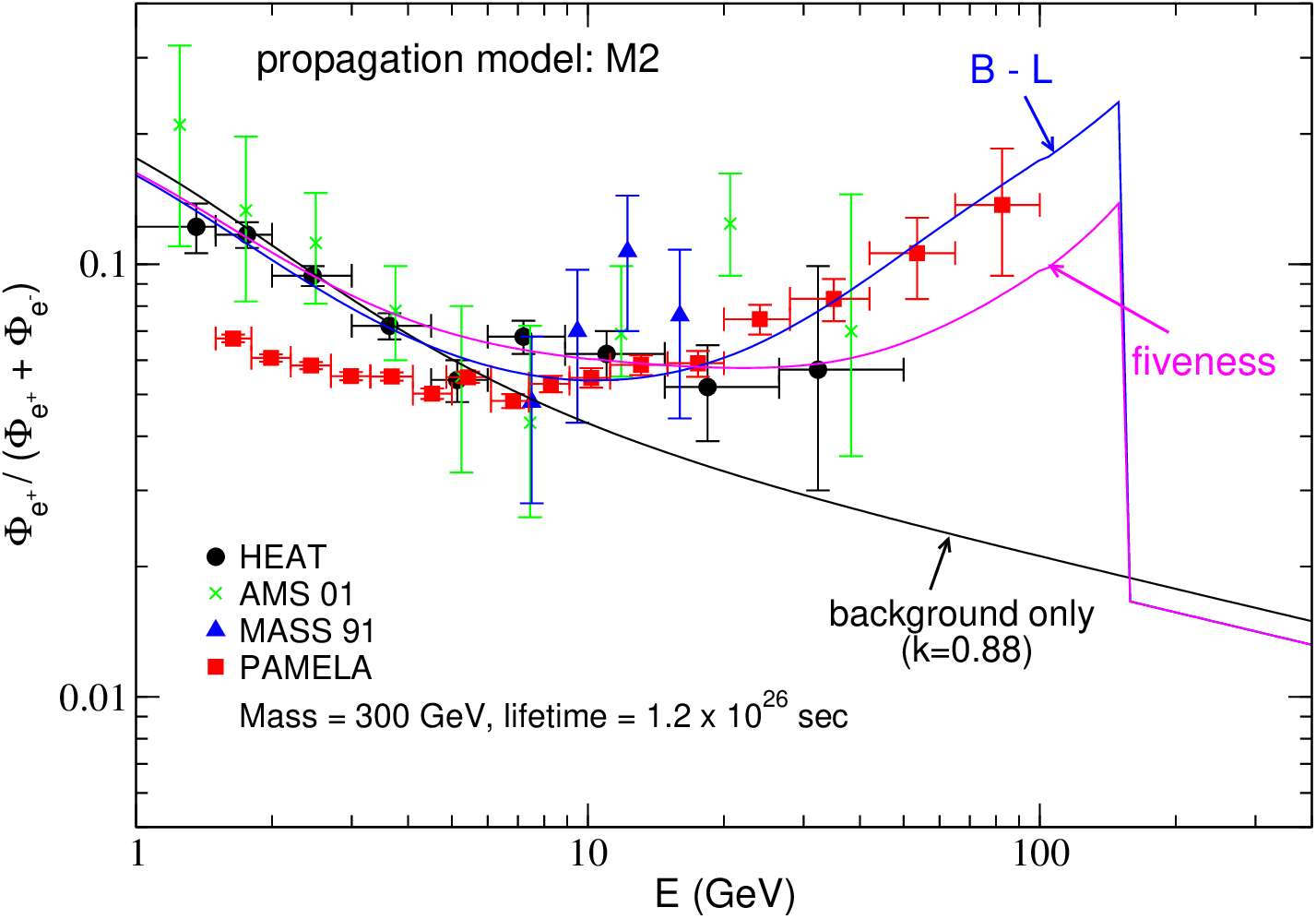}
\caption{The predicted positron fraction from $A_H$ decay via the kinetic mixing with $U(1)_{B-L}$ (blue line) and $U(1)_5$ (magenta line) using  the M2 propagation model,
compared with the experimental data, including the recent PAMELA results.}
\label{fig:positron}
\end{figure}

In our numerical calculations we set $m = 300$\,GeV and the lifetime $\tau = 1.2\times10^{26}$ seconds, and we use 
 the so-called  M2 diffusion model that are consistent with the Boron to Carbon ratio (B/C) and 
produce the minimal positron flux~\cite{Delahaye:2007fr,Ibarra:2007wg}. 
In Fig.~\ref{fig:positron} (blue line), we show the predicted positron fraction together 
with the recent PAMELA data and other experiments.
The prediction of our model fits very well with the excess reported by PAMELA.
The positron fraction steeply increases from $E \sim 10$\,GeV and drops off sharply
at \textbf{$E=m/2 = 150$}GeV, which is  mainly caused by the contribution of $e^{+}$
directly produced by the $A_H$ decay.
Such a drop-off can be checked by the upcoming PAMELA data in the higher energy region. 
 For the MED and M1 diffusion models~\cite{Delahaye:2007fr}, 
the positron fraction becomes slightly softer in the low energy (around several tens GeV), 
while the difference is negligible in the high end around the peak. 
We mention here that, with these new data of positron from PAMELA~\cite{Adriani:2008zr}, the background 
estimation may be different from what we adopted here~\cite{Moskalenko:1997gh,Baltz:1998xv}. However, since the 
signal of positron from decay of $A_H$ is negligible in the low energy and is important in the high energy region,
 i.e. $E\gtrsim 10$ GeV, we expect that our model will still be able to explain the excess even with such a new 
background estimation that fits the low energy data better.

\begin{figure}[t]
\includegraphics[scale=0.3]{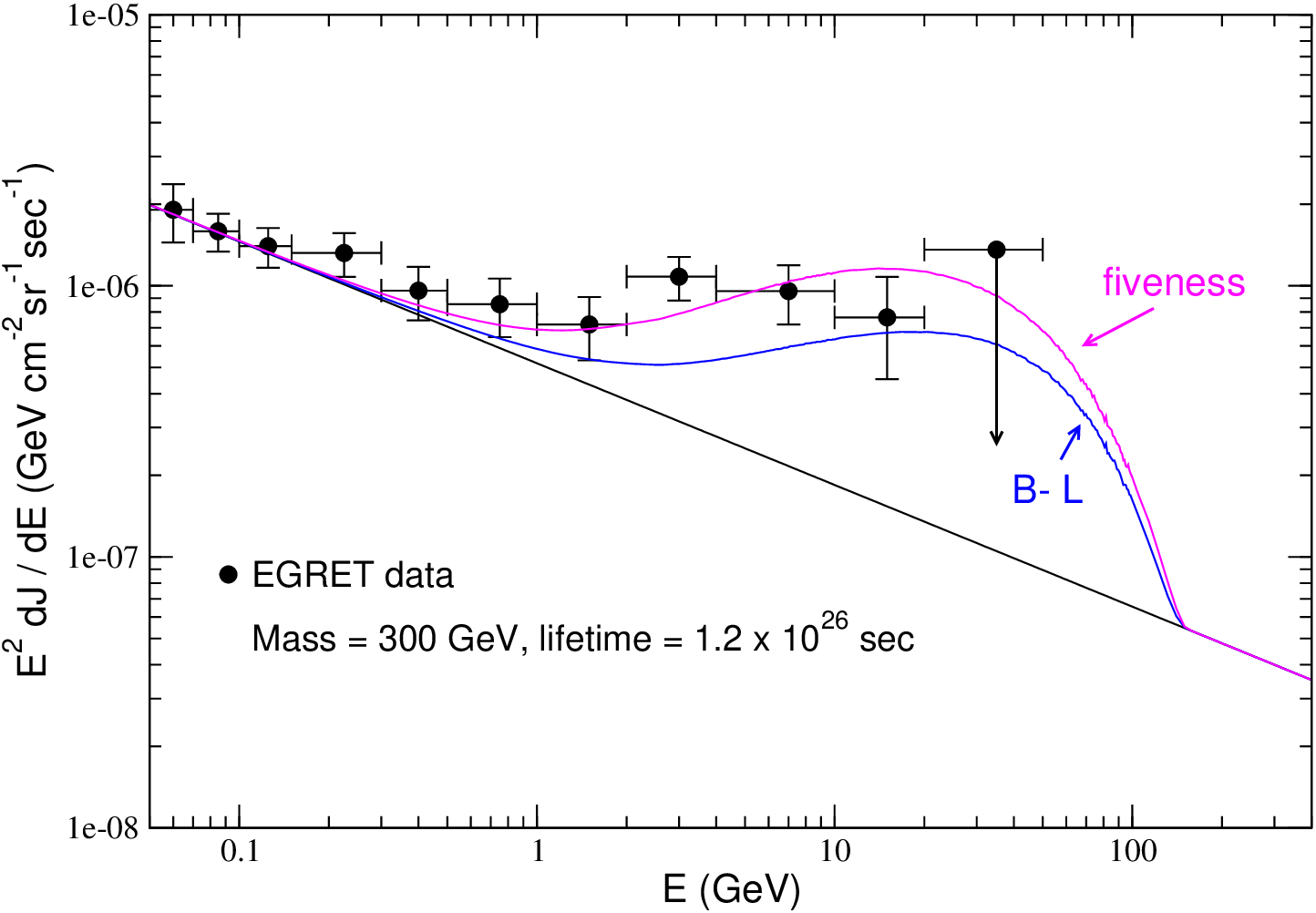}
\caption{The predicted gamma-ray flux of $A_H$ decay via the kinetic mixing with $U(1)_{B-L}$ (blue line) 
and $U(1)_5$ (magenta line), compared with the EGRET data}
\label{fig:gamma}
\end{figure}

The gamma-rays are mainly produced by the $\pi^{0}$ generated in the QCD hadronization process, since quark
pairs are produced from the decay of $A_H$.
In Fig.~\ref{fig:gamma} (blue line), we plot the gamma-rays together with the EGRET data. 
The gamma-ray flux peak 
at $E\sim 20$ GeV, and ends at $E = m/2$, which reflects the mass of decaying dark matter. 
With no surprise, we see
that the excess observed by EGRET may also be explained by the decay of $A_H$.  

\begin{figure}[t]
\includegraphics[scale=0.33]{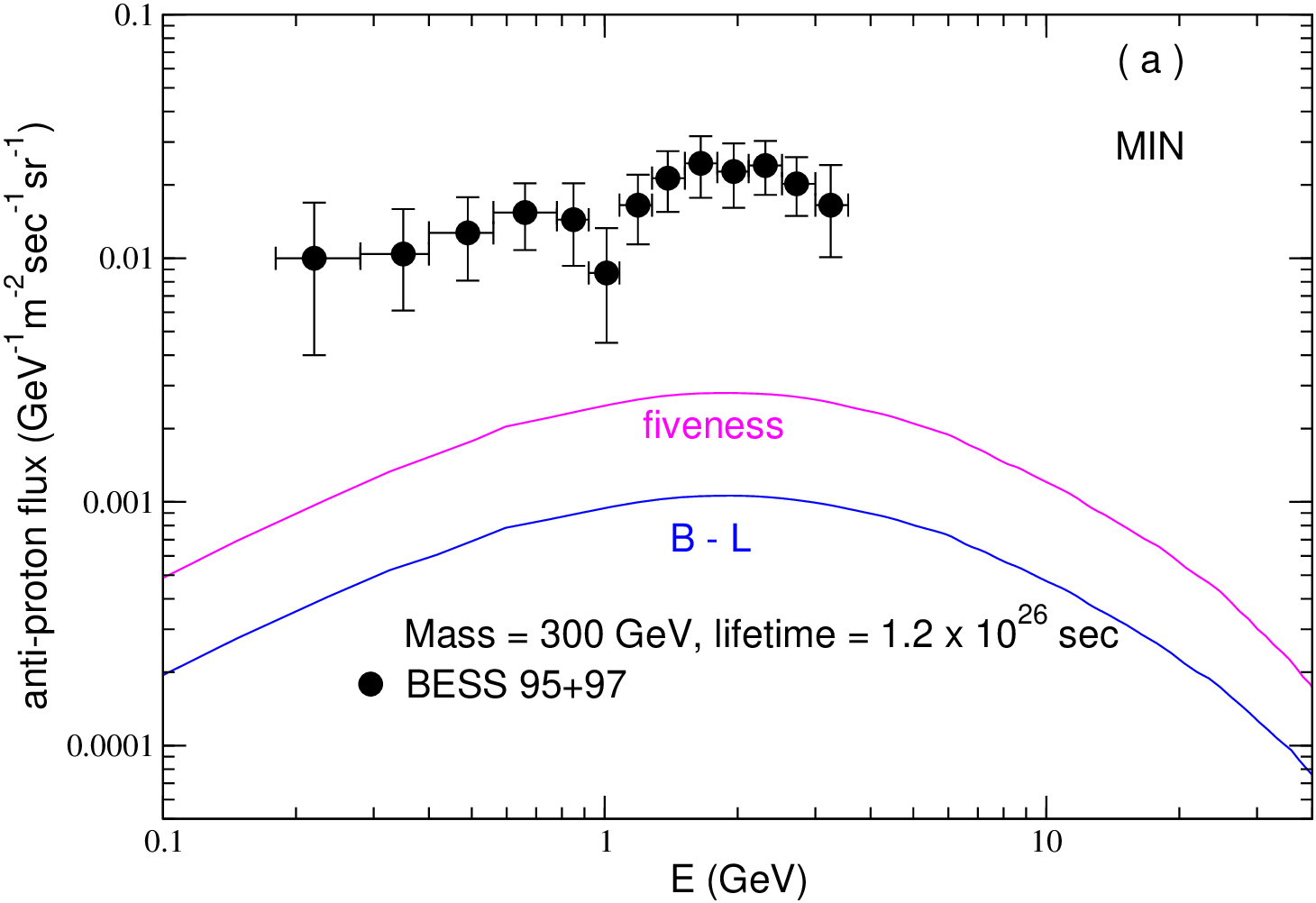}\ \includegraphics [scale=0.33]{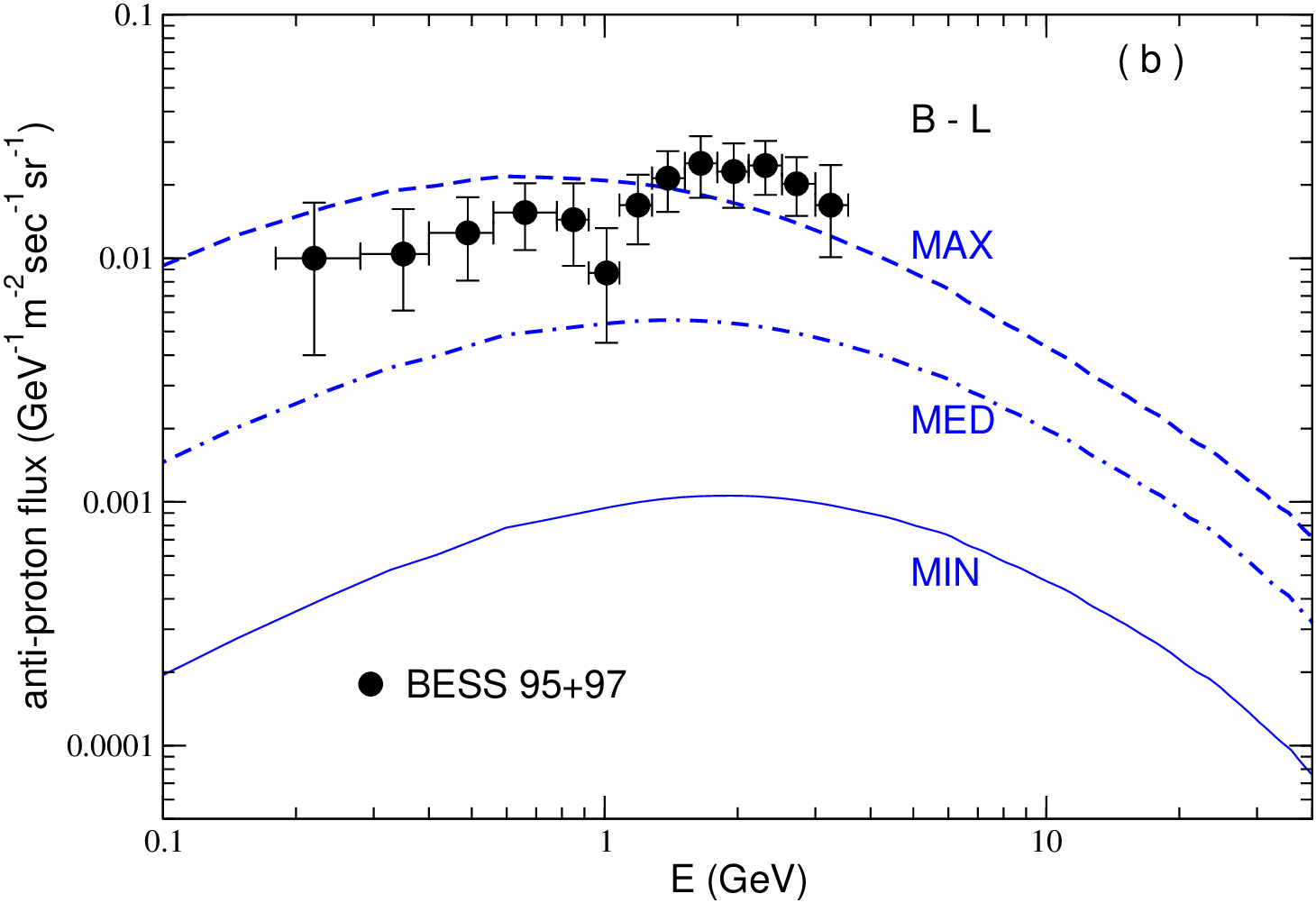}
\caption{(a) Antiproton flux from the $A_H$ decay via the kinetic mixing with $U(1)_{B-L}$ (blue line) 
and $U(1)_5$ (magenta line), compared with the BESS data. (b) Predicted antiproton fluxes 
from the $A_H$ in the $U(1)_{B-L}$ case, using different propagation models.
Note that the secondary antiproton contributions are not included.}
\label{fig:antip}
\end{figure}

Finally we show in Fig.~\ref{fig:antip} our predicted contribution to the antiproton flux (blue line).
In Fig.~\ref{fig:antip}, we have not included the prediction on the secondary antiproton flux,
which should explain the BESS data~\cite{Orito:1999re}.
Importantly, the predicted contribution to the antiproton flux from the $A_H$ decay
is smaller than the observed one by more than one order of magnitude, if the MIN propagation model~\cite{Donato:2003xg} is adopted, 
as shown in Fig.~\ref{fig:antip} (a). 
The suppression in the antiproton flux is particularly crucial,
because the recent data from PAMELA are consistent with the previous BESS result, which suggested that 
the secondary
production dominates the observed antiproton flux.
Furthermore, too many antiprotons tend to be generated as a by-product when
we require the dark matter annihilation/decay to account for the positron 
excess~\cite{Ibarra:2007wg,Cirelli:2008pk}.
Of course, the predicted antiproton flux still has a
large uncertainty mainly due to our poor understanding of the cosmic-ray
propagation inside our galaxy. As we can see in Fig.~\ref{fig:antip} (b), for different 
propagation models (MED and MAX), the antiproton flux from decaying $A_H$ can be enhanced by about 
two orders of magnitude. 
However, our scenario can still be consistent with the 
observed antiproton flux.\footnote{We notice that, the estimated background secondary antiproton flux drops quickly as the energy increases,
    and that the signal for the MED diffusion model seems to exceed the PAMELA data on the antiproton to proton ratio in the high energy
    region. However, our scenario is consistent with the PAMELA data in the high end if the MIN propagation model is adopted. }

\section{Discussion and Conclusions}
\label{sec:4} 
So far we have focused on the case that the $U(1)_m$ gauge symmetry in the bulk 
is identified with the $U(1)_{B-L}$ symmetry, but it is also possible to consider another
anomaly-free charge assignments given by a linear combination 
of the $B-L$ and the hypercharge $Y$. However, as we can check easily,
the hadronic decay branching ratio, which is a measure for the antiproton flux, becomes
smaller as the contribution of the hypercharge is suppressed. In this respect,
our choice of the $U(1)_{B-L}$ is well motivated by observation, since the
antiproton flux is naturally suppressed enough to be consistent with the
PAMELA data. For comparison, let us consider a $U(1)_5$, so-called ``fiveness", 
instead of $U(1)_{B-L}$. The charge $Q_5$ of the $U(1)_5$ is given by
the following linear combination of the charges under $U(1)_{B-L}$ and $U(1)_Y$~\cite{Fujii:2002mb}:
\beq
\frac{2 \sqrt{10}}{5} Q_5 \;=\; \frac{4}{5}\,Y - (B-L).
\eeq
We have similarly estimated the
spectra for the positron fraction, gamma-rays, and antiprotons 
when the $U(1)_m$ in the bulk is identified with $U(1)_5$. In order to compare with the $U(1)_{B-L}$ case, 
we take the same parameter sets, i.e. $m = 300$ GeV and $\tau =1.2\times 10^{26}$ sec. The decay branching ratios of 
$A_H$ decay are roughly 4.5\%, 18.3\%, 6.5\%, 4.8\% and 2.3\% for up-, down-type quark, 
charged lepton, neutrino and W-boson pairs, respectively.
We show the results of cosmic rays as the magenta lines in 
Figs.~\ref{fig:positron}, \ref{fig:gamma} and \ref{fig:antip}.
The distribution of positron fraction in fiveness case seems flatter compared to the B-L case and 
the inclination of the turnup is softer. This is due to the fact that positrons generated from the decay of
hadrons (mainly from $\pi^+$) become more important and the branching ratio of $e^+ e^-$ mode decreases simultaneously.
The fiveness model predicts a larger excess in the gamma-ray flux, since more $\pi^0$s are produced. Finally, 
we can see from Fig.~\ref{fig:antip} that the antiproton flux is enhanced for the $U(1)_5$
compared to the case of $U(1)_{B-L}$ due to a larger total branching ratio of quark pairs.  
Note that, in order to explain the PAMELA positron fraction data in the high energy region, we need to adopt a shorter lifetime 
for fiveness model compared to the
case of $B-L$. Accordingly, the gamma-ray and antiproton fluxes increase, 
which may result in a tension with observed data and make the fiveness model being slightly disfavored. 
In our previous work~\cite{Chen:2008yi} we considered a scenario that the hidden $U(1)_H$ gauge boson
mixes with the $U(1)_Y$. The predicted spectra in this case are somewhat between the $B-L$
and fiveness cases.

Let us comment on the production of the hidden gauge boson in the  
early Universe.
Although the couplings of $A_H$ to the SM particles are extremely  
suppressed,
we can generate a right abundance of $A_H$ from thermal
scatterings as follows. For  the reheating temperature about
$10^{15}$\,GeV, the $B-L$ gauge bosons will be in thermal equilibrium,
and the $A_H$ can be produced through
a non-renormalizable coupling such as
\beq
{\cal L} \simeq \frac{\kappa}{M_*^4} F^{(H)2} F^{(B)2},
\eeq
where $M_*$ denotes the five-dimensional Planck scale (see the  
footnote (1)) and
  $\kappa$ is a numerical coefficient of order unity.
The presence of such non-renormalizable operator on the hidden brane
is natural since it is allowed by the gauge symmetries.
The abundance of $A_H$ produced via the operator is roughly estimated by
\beq
\Omega_{AH} h^2 = {\cal O}(0.1)
  \lrfp{\kappa}{0.03}{2} \lrf{m}{300{\rm GeV}} \lrfp{T_R}{10^{15}{\rm  
GeV}}{3} \lrfp{M_*}{M_P/10}{-4},
\eeq
where $T_R$ is the reheating temperature.
Note that the required reheating temperature is smaller than the cut-off scale $M_*$.
Also non-thermal production of $A_H$ by e.g.  the
inflaton decay~\cite{Endo:2006qk,Endo:2006tf,Endo:2007ih} should work  
as well.

In this letter, we propose a scenario that a hidden gauge boson
constitutes the dark matter of the Universe and decays into the SM particles
via the kinetic mixing with the $U(1)_{B-L}$ gauge field in the bulk.
Interestingly, our model can account for the steep rise in the positron 
fraction reported by PAMELA as well as the gamma-ray excess seen 
by EGRET, while avoiding the constraint on the antiproton flux by PAMELA
and other experiments, if the M2 propagation model is adopted, due to the smallness of quark's quantum number under the gauged $U(1)_{B-L}$. 
Moreover, 
the very small decay rate of the hidden $A_H$ gauge boson dark matter is realized
naturally by  the hierarchy between weak scale and the large $B-L$ breaking scale which 
is about the GUT scale as suggested by the neutrino masses.

\begin{acknowledgments}
This work was supported by World Premier International Research Center
Initiative (WPI Initiative), MEXT, Japan. 
\end{acknowledgments}

\end{document}